\def\BibTeX{{\rm B\kern-.05em{\sc i\kern-.025em b}\kern-.08em
    T\kern-.1667em\lower.7ex\hbox{E}\kern-.125emX}}
\documentclass[12pt]{iopart}

\usepackage{iopams}  
\usepackage{cite}
\usepackage{amssymb,amsfonts}
\usepackage{algorithmic}
\usepackage{graphicx}
\usepackage{textcomp}
\usepackage{caption}
\usepackage{algorithm}
\usepackage{algorithmic}
\usepackage{soul}
\usepackage[utf8]{inputenc}
\usepackage[margin=1in]{geometry} 
\usepackage{booktabs}
\usepackage{multirow}
\usepackage{siunitx}

\begin{document}

\title[Motion Prediction inside Image Sensors]{A Retina-Inspired Pathway to Real-Time Motion Prediction inside Image Sensors for Extreme-Edge Intelligence}

\author{${}^{1}$Subhradip Chakraborty, ${}^{2}$Shay Snyder, ${}^{1}$Md Abdullah-Al Kaiser, ${}^{2}$Maryam Parsa, ${}^{3}$Gregory Schwartz, ${}^{1}$Akhilesh R. Jaiswal }

\address{${}^{1}$Department of Electrical and Computer Engineering, University of Wisconsin Madison,
    ${}^{2}$Department of Electrical and Computer Engineering, George Mason University,
    \\${}^{3}$Feinberg School of Medicine, Northwestern University}
\ead{chakrabort42@wisc.edu, akhilesh.jaiswal@wisc.edu}
\vspace{10 pt}
\begin{indented}
\item[]April 2025
\end{indented}

\begin{abstract}

The ability to predict motion in real time is fundamental to many maneuvering activities in animals, particularly those critical for survival, such as attack and escape responses. Given its significance, it is no surprise that motion prediction in animals begins in the retina. Similarly, autonomous systems utilizing computer vision could greatly benefit from the capability to predict motion in real time. Therefore, for computer vision applications, motion prediction should be integrated directly at the camera pixel level. 
Towards that end, we present a retina-inspired neuromorphic framework capable of performing real-time, energy-efficient MP directly within camera pixels. Our hardware-algorithm framework, implemented using GlobalFoundries' 22nm FDSOI technology, integrates key retinal MP compute blocks, including a biphasic filter, spike adder, nonlinear circuit, and a 2D array for multi-directional motion prediction. Additionally, integrating the sensor and MP compute die using a 3D Cu-Cu hybrid bonding approach improves design compactness by minimizing area usage and simplifying routing complexity. Validated on real-world object stimuli, the model delivers efficient, low-latency MP for decision-making scenarios reliant on predictive visual computation, while consuming only 18.56 pJ/MP in our mixed-signal hardware implementation. 
\end{abstract}


\section{Introduction}
Animal eyes are marvels of evolution, each uniquely adapted to meet ecological demands and enhance species survivability \cite{LAND2005R319}. Once thought of as mere organs for detecting and filtering light, eyes are now recognized as sophisticated structures that encode and process a vast range of visual information. This specialization is particularly pronounced in the retina, a peripheral part of the central nervous system containing parallel circuits that operate to extract distinct visual features ~\cite{eggers2011multiple}. Unlike biological eyes, engineered `eyes'—such as image sensors in machine vision—are rigid and lack computational capabilities or intelligence \cite{el2005cmos}. One critical feature the retina computes is motion prediction (MP), which allows animals to predict the future position of an object from its past trajectory. \cite{schwartz2021retinal}  This ability to anticipate motion is crucial for survival, allowing animals to react quickly to dynamic environmental changes.

Prediction is challenging because most environmental data lack predictive value, making it difficult to distinguish between predictive and non-predictive motion \cite{Kuo2016}. Since much of the environmental data do not aid in accurate forecasting, it becomes hard to identify movements or changes that genuinely contribute to prediction from those that do not. Thus, to use this capacity effectively, neural circuits prioritize predictive information, discarding non-predictive data during the encoding process. \cite{Berry} proposed an neuroscience MP model that relies on gain control. When a stimulus enters the edge of a retinal ganglion cell’s (RGC) receptive field, it triggers photoreceptor and bipolar cells, while also activating gain control mechanisms that reduce the light response volume. Although it is simple and effective, it doesn’t account for varying motion speeds. To address this issue of different stimulus velocities, later neuroscience research describes a nonlinear interaction through electrical and chemical synapses via gap junctions \cite{Trenholm2013}, which aids in MP. 

\begin{figure}[t]
    \centering
    \fbox{\includegraphics[width=0.9\textwidth]{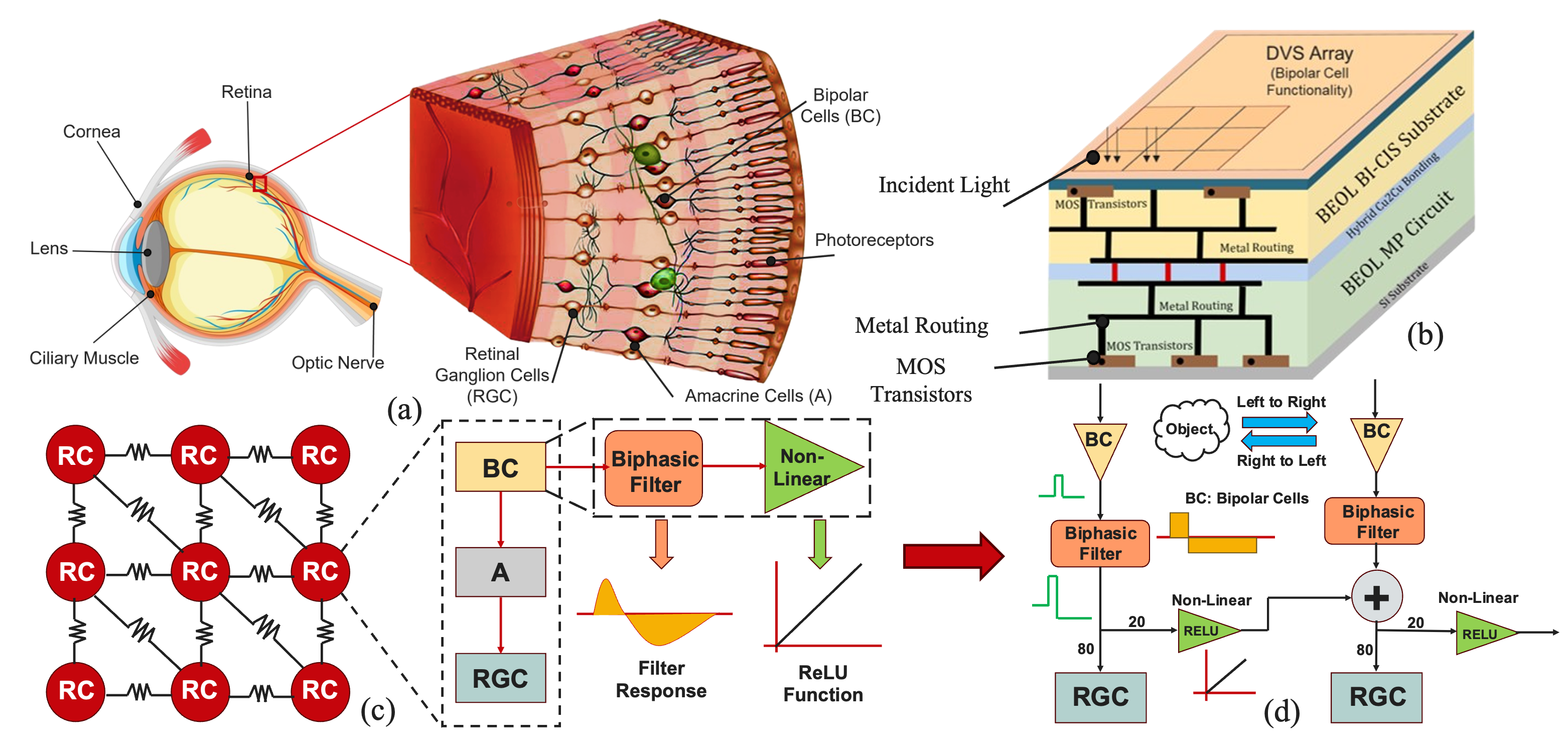}}
    \caption{(a) Representation of biological retina; (b) Proposed 3D integration camera with computational element; (c) Representation of connected retinal cells; (d) MP architecture and mechanism for a single direction motion.}
    \label{fig:intro}
\end{figure}

In computer vision applications, MP finds applicability in tasks such as object tracking, motion estimation, and scene understanding, where CMOS image sensors (CIS) combined with deep neural network algorithms are commonly used~\cite{1438751}. Previous work addressed motion tracking on-chip through hardware-software co-design, using traditional algorithms categorized into temporal difference methods (detecting changes in pixel intensities across frames)~\cite{4317686}, correlation methods (calculating the product of a pixel's intensity with a neighboring delayed intensity)~\cite{176627}, and cluster-based methods (grouping motion events by specific criteria)~\cite{6055400}. Although~\cite{6572366} proposes a hardware-friendly algorithm which performs motion tracking but it does not account for motion prediction. These works present challenges for on-chip implementation of motion tracking due to high power consumption and latency, and they do not account for MP.

To implement these above motion tracking algorithms on-chip while accounting for MP, efficient hardware architectures are required. Although efforts to mimic retinal circuits using solid-state technology date back to the 1980s~\cite{MEAD198891}, interest in integrating bio-inspired functionalities into electronic image sensors grew with the advent of neuromorphic sensors ~\cite{liu2010neuromorphic}, ~\cite{10.1063/5.0016485}. Recent work demonstrates retinal features like differential motion detection~\cite{6291964} and motion tracking~\cite{Li2024}, achieved through electronics and 2D materials~\cite{peng2024multifunctional}.
These works primarily focus on detecting and tracking past movement but cannot predict future motion. Additionally, some of them use 2D materials, which necessitate significant alterations to the existing CMOS foundry and face scaling and variability challenges.
Some studies have mimicked retinal characteristics using complex CMOS based biphasic circuits \cite{han2024design}, which, however, consume significant power and occupy substantial area (due to power hungry amplifier). Notably, these approaches focus on replicating the detailed electrochemical behavior of MP using mixed-signal circuits, rather than on image sensor or camera technology using retinal computation capabilities.

In this work, we propose a retina-inspired neuromorphic vision sensor that leverages a hardware-algorithm framework based on retinal neuroscience insights to perform pixel-level MP. Rather than simply replicating retinal functions, our approach develops a model, an algorithmic workflow, and a dedicated hardware architecture that seamlessly integrates the MP computational capabilities of the retina with those of a CMOS image sensor. To the best of our knowledge, this is the first work to present a retina-inspired neuromorphic camera for MP. In summary, the key contributions of this paper include:

\begin{enumerate}
    \item We identify and select only the essential critical features from the broad set of characteristics exhibited by biological retinal cells that are necessary to implement motion prediction (MP) functionality. This feature-specific approach minimizes design complexity, reduces energy consumption, and reduces area overhead, allowing integration of the proposed MP circuits into CMOS image sensors.
    \item We develop an algorithmic model inspired by neurological processes to enable precise MP at the pixel level, providing real-time prediction capabilities.
    \item We propose both a mixed-signal and a fully digital design for MP using GlobalFoundries' 22nm FDSOI technology. The mixed-signal design emulates essential retinal functions for MP and incorporates key compute elements, including a biphasic filter, signal divider, spike adder, and a non-linear circuit. It also features a 2D retinal circuit array capable of performing MP in all four directions. The fully digital implementation of the 2D MP architecture offers more flexible scaling characteristics using automated design tools.
    \item We proposed a 3D integration approach using Cu-Cu hybrid bonding, featuring one hybrid bond per pixel. This method reduces both area overhead and routing complexity, facilitating a more compact design. We validated the functionality of our MP model on real-world stimulus~\cite{ QIN2024106502}, demonstrating its practical implementation and pixel-level operating efficiency.
\end{enumerate}

\begingroup
\renewcommand{\arraystretch}{1.2}
\begin{table}[b]
\begin{center}
\caption{The mathematical symbols used to describe our simulated motion prediction framework.} 
\label{tab:symbol-table}  
\begin{tabular}{c|c}
\textbf{Symbol}        & \textbf{Meaning}                  \\ \hline
$k_x$                  & the number of RGCs along the x-axis \\
$k_y$                  & the number of RGCs along the y-axis \\
$k_s$                  & the receptive field of each RGC     \\
$k_o$                  & the percent overlap of each RGC     \\
$s$                    & kernel stride in pixels             \\
$T$                    & the number of discrete time bins    \\
$f_t$                  & time window of biphasic filter      \\
$\text{\textit{GJ}}_s$          & gap junction scaler               \\
$\text{\textit{GJ}}_{\text{\textit{nf}}}$       & gap junction normalization factor \\
$\gamma$                        & number of neighbors per RGC       \\
\end{tabular}
\end{center}
\vspace{-12pt}
\end{table}
\endgroup

The rest of the paper is organized as follows: Section \ref{sec_neuroMP} explains the mechanics of motion prediction in retina. Section \ref{sec_algoMP} and \ref{sec_circuitMP} presents the proposed retina inspired neuromorphic camera algorithm and hardware design. Section \ref{sec_results} shows the simulation results of the architecture. Section \ref{sec_conclusion} concludes the paper.

\section{Motion Prediction Circuit in the Retina} \label{sec_neuroMP}

In the vertebrate retina, photoreceptors form the initial layer, where they transduce light into analog voltage signals. These signals are then relayed to bipolar cells (BC) in the inner retina. There are approximately 15 types of BC \cite{Eggers2010-tr}, which in turn connect to around 60 types of amacrine cells \cite{Yan5177}, refining the signal for specialized computations. These refined signals are integrated by over 40 types of retinal ganglion cells (RGCs) \cite{GOETZ2022111040}, each with distinct receptive fields (RFs). RGCs transmit analog spikes that encode specific visual information, such as object motion, direction, orientation, and color contrast. Each RGC corresponds to a unique visual feature transmitted to the brain, enabling extraction of key features like motion prediction (MP) \cite{SCHWARTZ2021246}. Retinal cells send signals (chemical and electrical) to each other through gap junctions, which helps in performing MP. \cite{Kuo2016}

The implementation of MP, as shown in Figure \ref{fig:intro}(c), occurs in three main Stages. The first Stage involves BCs, which generate a signal upon luminance changes. Each BC includes a biphasic filter (BPF) followed by a nonlinear activation function. The BPF allows the forward (in the direction of motion) signal flow during the positive phase but prevents backward (opposite to the direction of motion) signal flow into the BCs during the negative or refractory phase, while the nonlinear circuit propagates the signal between BCs, arranged in a hexagonal grid. In the second Stage, a signal amplitude divider distributes the incoming signal—produced by convolving the spike with the BPF in a given ratio as, shown in Figure \ref{fig:intro}(d). The peak value is passed onto the first RGC. Finally, the third Stage adds the lower signal value (from the divider) after passing it through the non-linear circuit and combining it with the peak value of the second RGC. This design enables RGCs to generate amplified signals when a moving object activates a series of BCs along its path. The BCs in the motion path receive input from subsequent cells, amplifying their signal compared to those outside the path. 

In the following sections, we first implement the neuroscience model of MP as an algorithm. This approach facilitates the identification of key design parameters essential for hardware realization, while also offering a framework to assess the feasibility and efficiency of biological retinal circuits in extracting motion prediction behaviors. Unlike simplistic stimuli commonly employed in neuroscience experiments on surgically extracted retinas \cite{sekirnjak2006electrical}, our approach focuses on validating complex, real-world stimuli to better understand and replicate motion prediction behavior for technological use-cases \cite{ QIN2024106502}. 
Based on the neuroscience understanding of the MP circuit and algorithmic model, we then develop a mixed-signal CMOS-based MP hardware wherein spatio-temporal computations for MP are distributed in two 3D integrated chips via Cu-Cu hybrid bonds. 

\section{Algorithmic Implementation} \label{sec_algoMP}

In this section, we present the algorithmic model of MP and its correspondence with the neuroscience model discussed in Section 2. The implementation follows a structured multi-stage approach, where Stage 1 is captured in Algorithm 1 and Algorithm 2, handling the generation of BC signals and the application of a biphasic filter to process BC signals. Stage 2, implemented in Algorithm 3, models the interaction of gap junctions, facilitating lateral signal propagation and enhancing motion-related processing. Stage 3 is implemented in Algorithm 4 and Algorithm 5, describing the accumulation of higher amplitude signals during spike post-processing to refine the final representation. This structured approach ensures a clear transition from the neuroscience model to an algorithmic framework, laying the foundation for efficient hardware realization.

We implement a software defined implementation of our 3D MP circuit with PyTorch~\cite{paszke2019pytorchimperativestylehighperformance}, receiving BC signals (\textit{dvsSpikes}) as input and returning RGC activations \textit{MP} representing the likelihood of future motion. A detailed description of all parameters is shown in Table~\ref{tab:symbol-table}.

\begin{algorithm}
    \caption{Bipolar Cell Activations}
    \begin{algorithmic}
    \renewcommand{\algorithmicrequire}{\textbf{Require:}}
    \renewcommand{\algorithmicensure}{\textbf{Ensure:}}
    \REQUIRE dvsFrames, $k_s$, $k_o$
    \STATE bcKernel $\leftarrow$ buildGaussianKernel($k_s$)
    \STATE $s \leftarrow$ round($k_s * k_o$)
    \FOR{i = 0 to len(dvsFrames)}
        \STATE positiveSpikes $\leftarrow$ dvsFrames[i][+]
        \STATE BC[i] $\leftarrow$ applyKernel(bcKernel, s, positiveSpikes)
    \ENDFOR
    \RETURN \textit{BC}
    \end{algorithmic}
    \label{alg:bc-activations}
\end{algorithm}

\begin{algorithm}
    \caption{Bipolar Cell Non-Linearities}
    \begin{algorithmic}
    \renewcommand{\algorithmicrequire}{\textbf{Require:}}
    \renewcommand{\algorithmicensure}{\textbf{Ensure:}}
    \REQUIRE \textit{BC}, $f_t$
    \STATE $\text{\textit{filter}} \leftarrow \text{\textit{buildBiphasicFilter}}(f_t)$
    \FOR{t = 0 to len(\textit{BC})}
        \STATE $\text{\textit{BC\_NL\_POS}}[t] \leftarrow \text{applyKernel}(\text{\textit{BC}}[t], \text{filter})$ 
        \STATE $\text{\textit{BC\_NL\_NEG}}[t] \leftarrow \text{applyKernel}(\text{\textit{BC}}[t], -1 * \text{filter})$ 
    \ENDFOR
    \RETURN \textit{BC\_NL\_POS}, \textit{BC\_NL\_NEG}
    \end{algorithmic} 
    \label{alg:bc-nonlinearities}
\end{algorithm}

\textbf{Bipolar Cell (BC) Activations: } \textit{dvsSpikes} is a 4D matrix with the shape $(T, 2, H, W)$ where $T$ is the number of discrete time bins, $2$ is the polarity of each spike, $H$ is the height in pixels, and $W$ is the width in pixels. The MP circuit assumes that each DVS pixel is within a BC receptive field (RF) with an activation pattern approaching a normal distribution centered in the RF. We simulate this with a 2D convolution using a single input channel representing the incoming positive spikes and a single output channel with multiple features corresponding to the Gaussian responses from individual BCs. This process is shown in Algorithm~\ref{alg:bc-activations} utilizing three parameters and returning a matrix $BC$ representing the BC activations with shape ($T$, $k_x$, $k_y$). The size of the receptive fields is parameterized by the kernel size $k_s$ along with the stride that is calculated as $s \leftarrow$ round($k_s * k_o$) where $k_o = 0.5$ is the percent overlap between subsequent RFs.

\textbf{Bipolar Cell Non-linearites: } Next, the non-linear polarized BPFs are convolved along the time dimension of each BC. This process is shown in Algorithm~\ref{alg:bc-nonlinearities} accepting two arguments (\textit{BC} and $f_t$) and returning the positive and negative filtered outputs (\textit{BC\_NL\_POS} and \textit{BC\_NL\_NEG}).

\begin{algorithm}
\caption{Gap Junction Interactions}
\begin{algorithmic}[1]
\renewcommand{\algorithmicrequire}{\textbf{Require:}}
\renewcommand{\algorithmicensure}{\textbf{Continue}}
\REQUIRE $\text{\textit{BC\_NL}}, k_x, k_y, GJ_s$
\STATE $\text{\textit{GJ}} \in \mathbb{R}^{k_x \times k_y \times k_x \times k_y \times t}$
\STATE $value \leftarrow \text{\textit{BC\_NL}}[x, y, i] * \text{\textit{GJ}}_s$
\STATE $\mathcal{T} = \{0, 1, ..., T-1\}$
\STATE $X = \{0, 1, ..., k_x -1\}$
\STATE $Y = \{0, 1, ..., k_y -1\}$
\STATE $D = \{-1, 0, 1\}$
\FOR {$t, x, y, dx, dy \in \text{combos($\mathcal{T}, X, Y, D, D$)}$}
    \STATE $nx \gets x + dx$; $ny \gets y + dy$
    \IF {$\text{\textbf{not} } (0 \leq nx < k_x$ and $0 \leq ny < k_y)$}
        \STATE \textbf{skip iteration}
    \ENDIF
    \STATE $\textit{incoming} \gets \text{False}; \textit{outgoing} \gets \text{False}$
    \IF{($ny < y$ \AND ($nx \leq x$)) \OR ($ny==y$ \AND $nx<x$)}
        \STATE $\textit{incoming} \gets True$ 
    \ELSIF{($ny > y$ \AND $nx \geq x$) \OR ($ny==y$ \AND $nx>x$)}
        \STATE $\textit{outgoing} \gets True$
    \ENDIF

    \IF{$\textit{incoming}$}
        \STATE $GJ[nx, ny, x, y, t] \mathrel{+}= \text{value}$
    \ELSIF{$\textit{outgoing}$}
        \STATE $GJ[x, y, nx, ny, t] \mathrel{-}= \text{value}$
    \ENDIF
\ENDFOR
\RETURN \textit{GJ}
\end{algorithmic}
\label{alg:gj}
\end{algorithm}

\textbf{Gap Junction (GJ) Interactions: } Our MP circuit also includes support for localized interactions between the individual RGCs, with another procedure shown in Algorithm~\ref{alg:gj} that accepts 4 arguments (\textit{BC\_NL}, $k_x$, $k_y$, and $\text{\textit{GJ}}_s$) and returns a single 5D matrix \textit{GJ} representing the incoming and outgoing interactions across all GJ. The GJ normalizer $\text{\textit{GJ}}_s$ inhibits the influence of neighbor activity for each individual RGC. It is calculated as $\text{\textit{GJ}}_s = \text{\textit{GJ}}_{\text{\textit{nf}}} / \gamma$ where $\text{\textit{GJ}}_{\text{\textit{nf}}}$ is a real number between 0 and 1 and $\gamma$ is the number of neighbors. For all experiments, we specify $\text{\textit{GJ}}_{nf} = 0.5$ and $\gamma = 6$. We apply this function across both positive and negative non-linear activations (\textit{BC\_NL\_POS} and \textit{BC\_NL\_NEG}) resulting in two 5D matrices: \textit{BC\_NL\_POS\_GJ} and \textit{BC\_NL\_NEG\_GJ}.

\textbf{Gap Junction Interaction Accumulation: } With the GJ interactions calculated for both the positive and negative BC non-linearities, we accumulate the difference of the sum of incoming and outgoing interactions as shown in Algorithm~\ref{alg:gj-int}. We apply this process over both GJ matrices (\textit{BC\_NL\_POS\_GJ} and \textit{BC\_NL\_NEG\_GJ}) and both activation matrices (\textit{BC\_NL\_POS}, \textit{BC\_NL\_NEG}) resulting in two BC activation matrices with GJ interactions (\textit{POS\_GJ}, \textit{NEG\_GJ}).

\begin{algorithm}
    \caption{Accumulate Gap Junction Interactions}
    \begin{algorithmic}
    \renewcommand{\algorithmicrequire}{\textbf{Require:}}
    \renewcommand{\algorithmicensure}{\textbf{Ensure:}}
    \REQUIRE $\text{\textit{BC}}, \text{\textit{GJ}}, T, k_x, k_y$
    \STATE $\mathcal{T} = \{0, 1, ..., T-1\}$
    \STATE $X = \{0, 1, ..., k_x -1\}$
    \STATE $Y = \{0, 1, ..., k_y -1\}$
    \FOR{$t, x, y \in \text{combos}(\mathcal{T}, X, Y)$}
        \STATE $in \leftarrow \sum GJ[:, :, x, y, t]$
        \STATE $out \leftarrow \sum GJ[x, y, :, :, t]$
        \STATE $\text{\textit{BC}}[x, y, t] \mathrel{+}= in - out$
    \ENDFOR
    \RETURN \textit{BC}
    \end{algorithmic} 
    \label{alg:gj-int}
\end{algorithm}

\begin{algorithm}
    \caption{Post-Rectification}
    \begin{algorithmic}
    \renewcommand{\algorithmicrequire}{\textbf{Require:}}
    \renewcommand{\algorithmicensure}{\textbf{Ensure:}}
    \REQUIRE \textit{POS\_GJ}, \textit{NEG\_GJ}
    \STATE $\text{\textit{POS\_GJ}}[\text{\textit{POS\_GJ}} < 0] = 0$
    \STATE $\text{\textit{NEG\_GJ}}[\text{\textit{NEG\_GJ}} < 0] = 0$
    \STATE $\text{\textit{RGC}} = \text{\textit{POS\_GJ}} + \text{\textit{NEG\_GJ}}$
    \RETURN \textit{MP}
    \end{algorithmic} 
    \label{alg:post-rec}
\end{algorithm}

\textbf{Post Rectification: } Aggregating everything together, we remove all elements less than zero from both activation matrices and accumulate everything into the final RGC neural activity \textit{MP}. This process is shown procedurally in Algorithm~\ref{alg:post-rec}.

\section{Hardware Implementation} \label{sec_circuitMP}

This section presents the hardware architecture of MP and its correspondence with the neuroscience and algorithmic models discussed in Sections 2 and 3. The circuit implementation follows a structured multi-stage approach, mirroring the algorithmic framework. Stage 1, described in Section 2, is implemented using a Dynamic Vision Sensor (DVS) to generate BC spikes, corresponding to Algorithm 1. Algorithm 1 describes the use of a Gaussian filter, but we neglect this in hardware design as it does not affect the effectiveness of the MP circuit. Similar, approximation neglecting the Gaussian filtering has been used for other retinal features like object motion sensitivity as shown in \cite{yin2023iris}. Algorithm 2, which defines the biphasic filter, is realized through a combination of an MP block and a Motion-Selective (MS) filter, detailed later in this section. The 1D Motion Prediction subsection implements Stages 2 and 3, corresponding to Algorithms 3 and 4, where amplified spikes are generated based on object motion. Additionally, the hardware also generates predictive spikes before the object reaches the next BC, enabling the implementation of real-time MP on-chip. Finally, the 2D MP subsection extends the 1D architecture, scaling it to a hexagonal grid while incorporating a readout circuit for spike detection. This structured approach ensures that the hardware accurately replicates both the neuroscience and algorithmic models, maintaining fidelity to the underlying biological processes.

\subsection{Dynamic Vision Sensor (DVS) - Bipolar Cells (BC) Activation} \label{sec_dvs}

The dynamic vision sensor (DVS) camera pixel emulates retinal bipolar cells by employing a contrast-sensitive mechanism that generates ON and OFF bipolar signals (BP) in response to scene contrast variations, as illustrated in Figure \ref{fig:MP1}(a) \cite{DVS_ref1, DVS_ref2}. The circuit comprises a logarithmic receptor, voltage buffer, difference amplifier, and thresholding components. The logarithmic photoreceptor converts incoming light into a logarithmic voltage output, which is then isolated by a source follower buffer (\si{\times 1}) to protect the sensitive pixel node (\si{V_{LOG}}) before being fed into the difference amplifier. A capacitive feedback difference amplifier \cite{DVS_ref3} asynchronously computes the voltage gradient corresponding to changes in light intensity. Finally, the output voltage of the difference amplifier (\si{V_C}) is processed by two thresholding circuits—one comparing contrast increases with a threshold and the other comparing contrast decreases with a threshold—to generate an ON/OFF bipolar signal. For our Mp circuit, detailed in subsection \ref{sec_mp1d}, we utilize bipolar signals from DVS pixels as input, regardless of their polarity.

\subsection{1D Motion Prediction - Gap Junction Interaction} \label{sec_mp1d}

\begin{figure}[t]
    \centering
    \fbox{\includegraphics[width=0.9\textwidth]{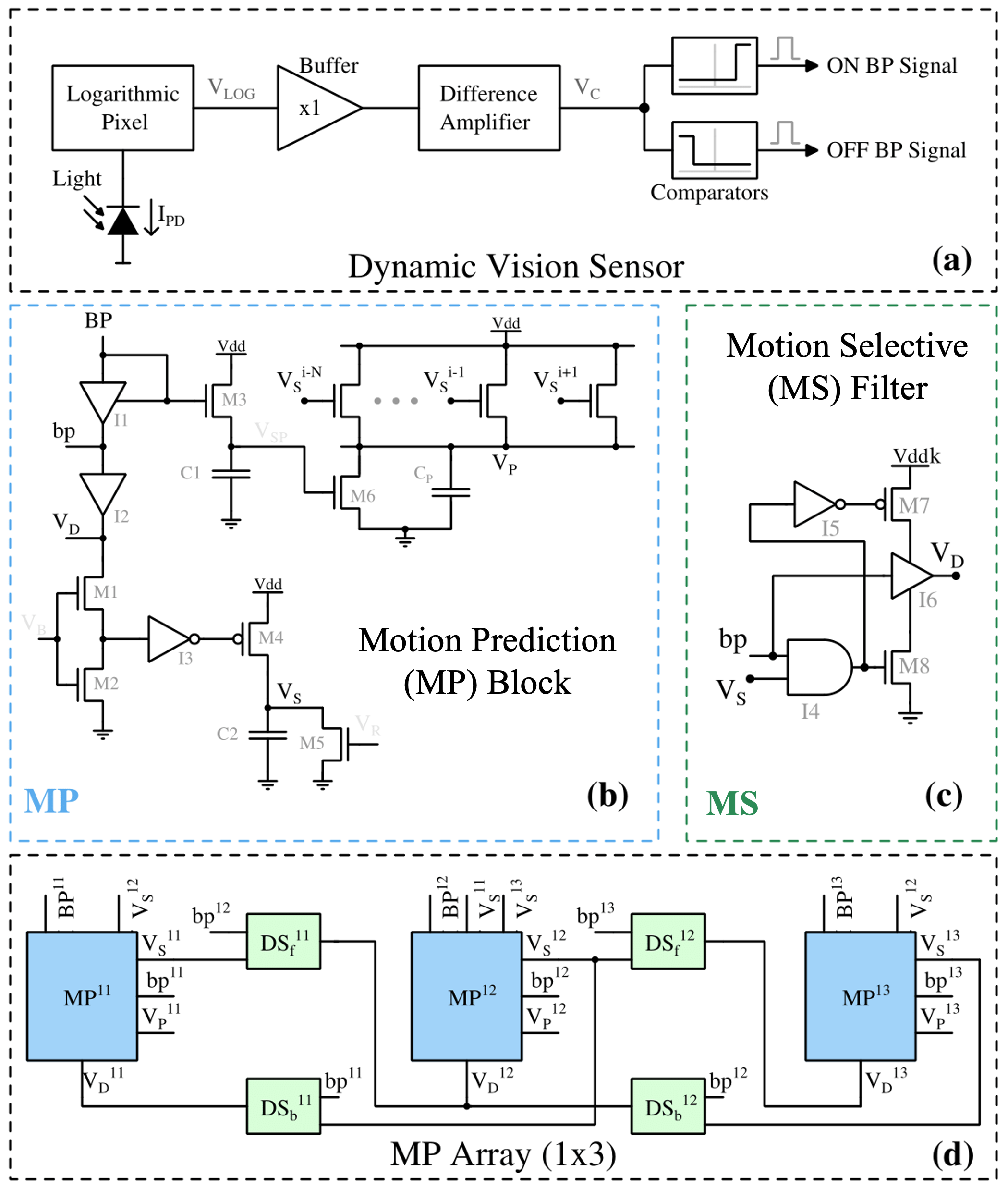}} 
    \caption{(a) DVS pixel circuit implementation generating ON and OFF bipolar signals (BP); (b) Proposed motion prediction (MP) circuit generating predictive spikes; (c) Motion-Selective (MS) circuit replicating the functionality of a biphasic filter; (d) 1\si{\times}3 MP array illustrating the connectivity between adjacent MP blocks.}
    \label{fig:MP1}
\end{figure}

Figure \ref{fig:MP1}(b) shows the schematic of the proposed 1D motion prediction (MP) block. The input bipolar signals (\si{BP}) from each DVS pass through a power-gated buffer (I1) and are sent to the bipolar node (\si{bp}) of the MP block, serving as the input for the MP circuit. This design minimizes power consumption by enabling signal propagation only when a bipolar signal is generated. Additionally, these buffers can operate at a lower supply voltage than the DVS sensor’s, allowing power and speed optimization of the compute circuit in advanced process nodes. Another buffer (I2) connects the \si{bp} node to \si{V_D}, propagating the bipolar signal when the object with predictive motion first activates the DVS pixels. 

The node \si{V_D}, which emulates the output functionality of MP RGC cells, is connected to a network of resistive dividers modeled using NFET transistors (M1 and M2) operating in the linear region. Next, an inverter-based comparator circuit (I3) with a fixed threshold processes the signal. The voltage at the \si{V_D} node is scaled by the divider network, mimicking the gap junction fraction, before being fed into the comparator (I3). When the scaled \si{V_D} voltage exceeds the comparator threshold, this information is stored by precharging the MOM capacitor (C2) via the M4 transistor, as shown in Figure \ref{fig:MP1}(b). This capacitor manages the time lag between the bipolar signals caused by changes in the object's speed by temporarily storing the signal value until the next bipolar signals arrive, ensuring reliable motion prediction.

The biphasic filter (BPF), as discussed in section \ref{sec_neuroMP}, operates in two distinct phases: a positive phase, where signals are allowed to pass, and a negative (or refractory) phase, where incoming signals are blocked. This filtering mechanism is crucial for detecting motion in a specific direction. For instance, when a stimulus moves from left to right, adjacent bipolar cells generate signals in response to the change in light intensity. The BPF ensures that only forward-moving signals (left-to-right) are transmitted, while suppressing any backward-moving signals (right-to-left).
The BPF is implemented using the Motion-Selective (MS) circuit as shown in Figure \ref{fig:MP1}(c), which determines motion direction by analyzing the present and past locations of spikes in surrounding cells. The storage node \si{V_S} holds information about past spikes of the adjacent cells, while the \si{bp} node represents the bipolar signal information of the current cell. If both spikes occur simultaneously, the MS circuit generates an amplified spike (similar to a spike adder), which is then sent to the \si{V_D} node of current cells. The amplified spike is created by using a supply voltage of \si{Vddk}, which is higher than the \si{Vdd} used in the MP circuit and through buffer I6.

The predictive node (\si{V_P}) is designed to identify pixels where an object is expected to move next while excluding locations it has already passed. It consists of a spike storage node (\si{V_{SP}}) that records spike occurrences at a given location and integrates predictive logic to anticipate motion. When adjacent pixels have previously generated spikes, their corresponding MP blocks store this information at node \si{V_S}, leading to the charging of capacitor (\si{C_P}). However, if a new \si{BP} signal is detected at the current pixel, indicating the object's presence, the \si{V_{SP}} node discharges the capacitor, dynamically clearing the predicted movement path. Let's analyze the process when an object moves \si{MP^{11}} to \si{MP^{12}} in a left-to-right direction, illustrating the step-by-step operation of the motion prediction.

\begin{itemize}
    \item When the object reaches \si{MP^{11}}, the spike storage node \si{V_{SP}^{11}} becomes charged, indicating its presence. Since \si{MP^{11}} is the first motion prediction (MP) block encountered, it does not receive any predictive signal \si{V_S^{12}} from its neighboring MP block \si{MP^{12}}. As a result:  
    \begin{itemize}
        \item The predictive node \si{V_P^{11}} remains at zero because the charged \si{V_{SP}^{11}} establishes a discharge path through the M6 transistor.  
        \item Meanwhile, \si{MP^{12}} receives the predictive signal \si{V_S^{11}} from \si{MP^{11}}, causing \si{V_P^{12}} to charge. This indicates the anticipated movement of the object toward \si{MP^{12}}.  
    \end{itemize}  

    \item As the object moves to \si{MP^{12}}, the generation of \si{BP^{12}} triggers the charging of the spike storage node \si{V_{SP}^{12}} via the M3 transistor. As a result:  
    \begin{itemize}
        \item The M6 transistor establishes a discharge path for the capacitor \si{C_P} within \si{MP^{12}}, resetting the predictive node \si{V_P^{12}} to zero.  
        \item \si{MP^{12}} transmits the predictive signal \si{V_S^{12}} to \si{MP^{11}}, but since \si{V_{SP}^{11}} retains its previous charge, it prevents the charging of \si{V_P^{11}}. This ensures that previously visited locations are excluded from anticipation.  
    \end{itemize}  

    \item Initially, when the object was at \si{MP^{11}}, the circuit anticipated \si{MP^{12}} as the next likely location. However, upon reaching \si{MP^{12}}, it does not predict \si{MP^{11}} again, as that location has already been visited.  

    \item The number of transistors connected to the predictive node \si{V_P} depends on the number of future locations to be anticipated. For example, if the system is designed to predict four positions ahead in each direction, a total of 16 transistors would be required.  
\end{itemize}

\subsection{2D Motion Prediction Array} \label{sec_mp2d}

\begin{figure}[t]
    \centering    
    \fbox{\includegraphics[width=0.9\textwidth]{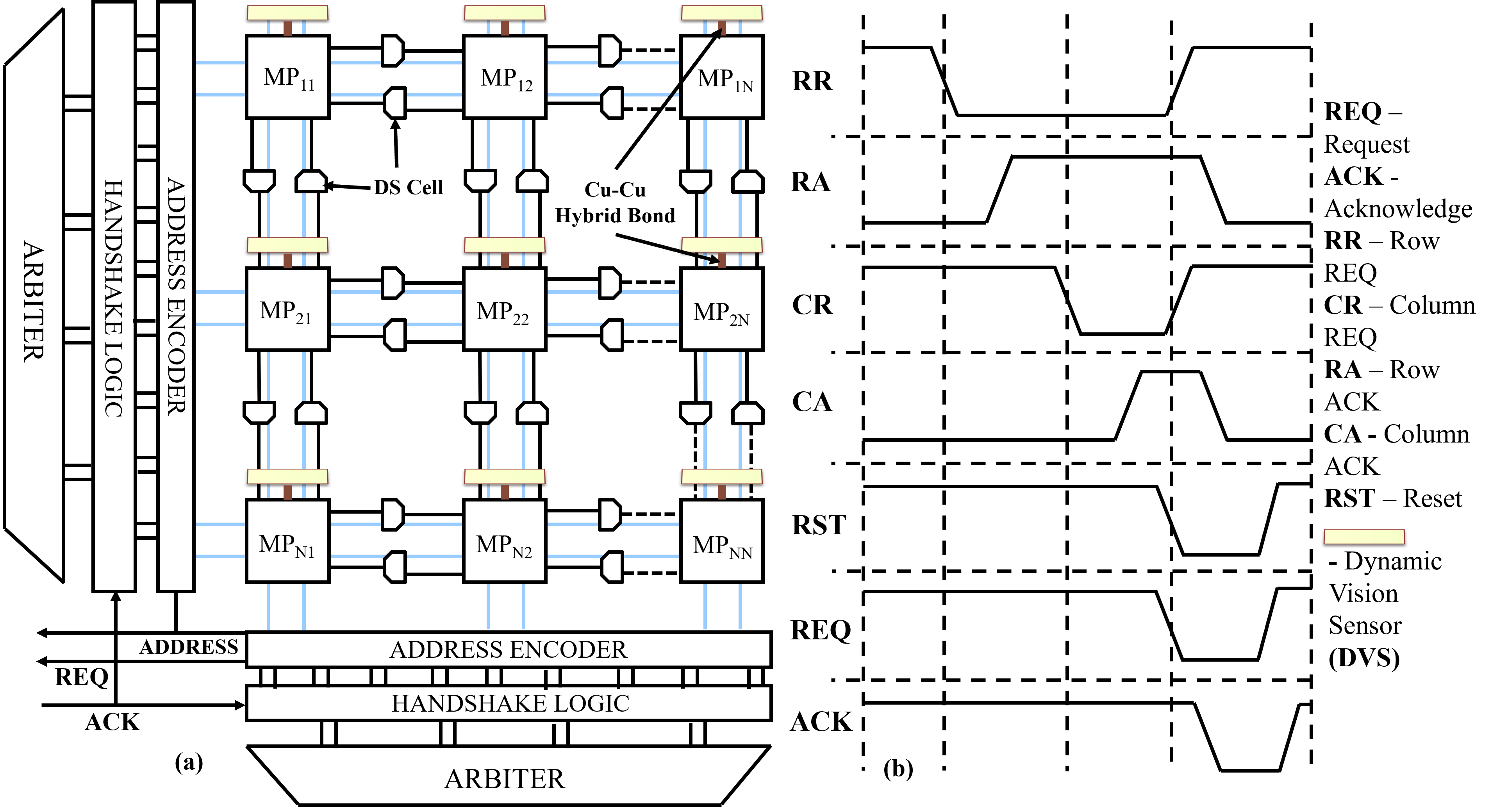}} 
    \caption{(a) Array-level Motion Prediction (MP) architecture, integrated with a conventional AER readout mechanism; (b) Timing diagram for the AER readout process for MP spikes.}
    \label{fig:MP2}
\end{figure}

The 1D array has two directional paths: a forward path for left-to-right motion and a backward path for right-to-left motion. During left-to-right movement, the forward path serves as the primary route for motion prediction (MP), while the backward path is crucial for right-to-left motion. Between two MP blocks, as shown in Figure \ref{fig:MP2}(a), there are two Motion-Selective (MS) blocks. In the case of a 2D MP array, a similar strategy is applied across columns to enable top-to-bottom and bottom-to-top motion prediction, with each MP block connected to four MS blocks to facilitate MP in all directions. When motion occurs from top to bottom or bottom to top, the RGC nodes (\si{V_D}) and the predictive voltage nodes (\si{V_P}) are charged and discharged according to the method described in Section \ref{sec_mp1d}.
Note, diagonal connections are excluded from the circuit design because, given the practical size difference between the object and the pixels, they would contribute to increased power consumption without enhancing prediction performance. The object is adequately large to be detected through horizontal and vertical connections with reasonably good performance. Moreover, the predictive level—defined by the number of surrounding cells the object may move to—can be extended by adding more NFETs in parallel at the (\si{V_P}) node circuit, enabling greater anticipation of potential movement. The predicted spikes can be transmitted using the standard Address Event Representation (AER) method, following a conventional timing diagram \cite{AER_ref1, AER_ref2}. The handshaking signals, including row request (RR), row acknowledge (RA), column request (CR), and column acknowledge (CA), are illustrated in Figure \ref{fig:MP2}.

\begin{figure}[t]
    \centering
    \fbox{\includegraphics[width=0.9\textwidth]{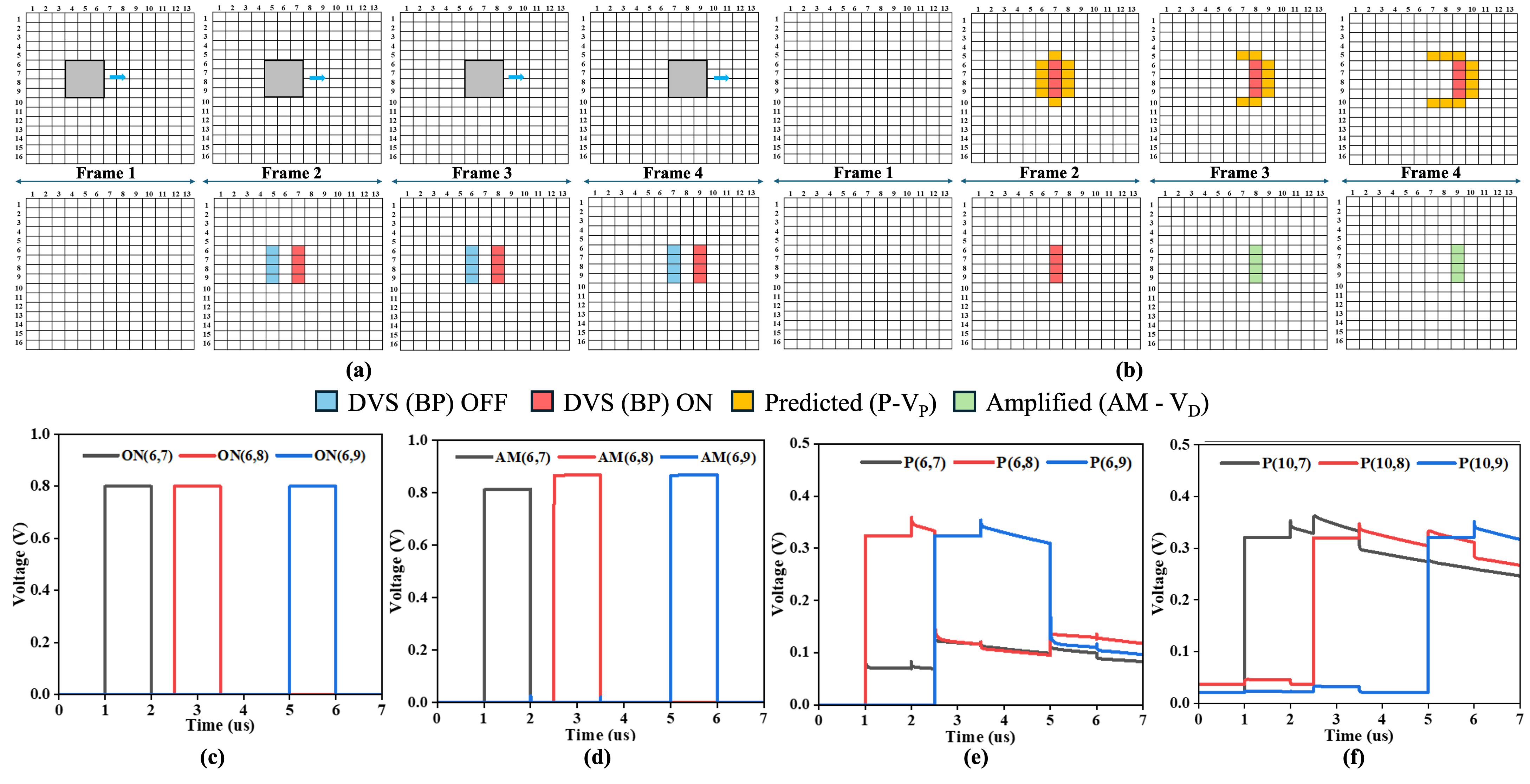}} 
    \caption{(a) (Top) The object's movement from left to right. (Bottom) DVS bipolar signals generated due to contrast changes. (b) (Top) Predictive node (P) spikes generated and (Bottom) amplified node (AM) spikes generated during left-to-right movement. (c) DVS bipolar signals for pixel locations (6,7), (6,8), and (6,9). (d) Corresponding amplified spikes. (e) and (f) Corresponding predictive spikes. Note, the amplified spike (AM) at the RGC node corresponds to the \si{V_D} node, while the predicted spike (P) corresponds to the \si{V_P} node.}
    \label{fig:mpcase}
\end{figure}

Figure \ref{fig:mpcase} demonstrates a one-level motion prediction mechanism, utilizing four NFETs to enable movement detection by one pixel in four possible directions. To illustrate the circuit's functionality step by step, consider an object moving from left to right, as shown in Figure \ref{fig:mpcase}. The amplified spike (AM) at the RGC node corresponds to the \si{V_D} node, while the predicted spike (P) corresponds to the \si{V_P} node. As the object transitions from frame 1 to frame 2, the DVS pixels generate bipolar signals due to contrast variations caused by the object's movement. In this example, \si{bp(6,7)}, \si{bp(7,7)}, and \si{bp(8,7)} produce bipolar signals shown in red in frame 2 of Figure \ref{fig:mpcase}(a) (an example bipolar signal for the pixel location (6,7) is shown in Figure \ref{fig:mpcase}(c) at 1 \si{\micro s}), which are then transmitted to their respective RGC nodes (\si{V_D}) and stored in the \si{V_S} nodes. Within this 2D grid representation, the first number in parentheses indicates the row index, while the second represents the column index. Since spikes are present at the \si{V_S} nodes of these cells, the surrounding cells in all four directions generate predictive spikes at their respective \si{V_P} nodes, as depicted with yellow in frame 2 (top) of Figure \ref{fig:mpcase}(b). For instance, the predictive node voltage at pixel location \si{(6,8)} goes high, as seen in Figure \ref{fig:mpcase}(e) at 1 \si{\micro s}. A similar behavior is observed in the predictive nodes of the surrounding cells. Notably, since the object appears in these pixels for the first time, the corresponding RGC nodes output bipolar-like signals (without amplification from the MS block), as observed in \si{AM(6,7)} shown in green in Figure \ref{fig:mpcase}(d) at 1 \si{\micro s}. A similar behavior will be observed at the RGC nodes of these cells, as shown in frame 2 (bottom) of Figure \ref{fig:mpcase}(b).

As the object transitions from frame 2 to frame 3, the DVS generates bipolar signals at pixel locations (6,8), (7,8), and (8,8), resulting in \si{bp(6,8)}, \si{bp(7,8)}, and \si{bp(8,8)} receiving these signals, as illustrated in frame 3 of Figure \ref{fig:mpcase}(a). Consequently, the predictive nodes of the surrounding pixels generate spikes due to updates in the \si{V_S} nodes at these locations. The spikes in the predictive nodes are depicted in frame 3 (top) of Figure \ref{fig:mpcase}(b), with an example spike at \si{P(6,9)} shown in Figure \ref{fig:mpcase}(e) at 2.5 \si{\micro s}. Notably, the predictive nodes of pixels previously occupied by the object are reset, as exemplified by the \si{P(6,8)} node decreasing, as seen in Figure \ref{fig:mpcase}(e) at 2.5 \si{\micro s}. Additionally, the RGC nodes corresponding to pixel locations (6,8), (7,8), and (8,8) receive amplified signals from the MS circuits, as observed in \si{AM(6,8)} in Figure \ref{fig:mpcase}(d) at 2.5 \si{\micro s}. As the object continues to move, the surrounding predictive nodes generate spikes, while all the pixels the object has already passed are reset. The paths that the object did not cross, but were predicted, show a slower decay, as seen in Figure \ref{fig:mpcase}(f), where pixel locations (10,7), (10,8), and (10,9) generated predicted spikes during the object's movement. Meanwhile, the RGC nodes along the object's path receive amplified spikes, confirming that the object is following one of the predicted paths.

\subsection{3D Integration}\label{sec_3dinteg}

Figure \ref{fig:intro}(b) presents a 3D heterogeneously integrated design of our proposed retina-inspired camera for motion prediction (MP) \cite{3D_integ1, 3D_integ2}. The system is composed of two distinct dies: (1) a backside-illuminated CMOS image sensor (BI-CIS) that accommodates the DVS pixels and bias circuitry, and (2) a bottom die incorporating the 2D motion prediction compute array, Motion-Selective circuit, AER readout circuit, and other peripheral components. A major advantage of this heterogeneous integration lies in the ability to fabricate the bottom die using an advanced technology node while keeping the DVS pixel array in a separate BI-CIS die, typically designed in a lagging process node. The adoption of 3D stacking significantly reduces routing complexity and enhances processing efficiency (due to short routing), while the integration of MOM capacitors above transistors optimizes area utilization without compromising pixel density. Moreover, the design employs fine-pitched hybrid Cu-Cu bonding \cite{Cu2Cu_pitch}, enabling compact and efficient interconnects. Each DVS pixel transmits its bipolar signal to the MP compute circuit via a single Cu-Cu hybrid bond. Given that DVS pixels are relatively large \cite{DVS_ref1}, the bonding pitch and MP circuit can be precisely aligned with the top-layer DVS pixels. Furthermore, 3D integration enhances transmission energy efficiency over traditional 2D integration by minimizing interconnect length \cite{tri_design}.

\begin{algorithm}
\caption{Digital Flow of Motion Prediction}
\begin{algorithmic}[1]
\renewcommand{\algorithmicrequire}{\textbf{Require:}}
\renewcommand{\algorithmicensure}{\textbf{Ensure:}}
\REQUIRE $DVS[N][N]$
\ENSURE $amplified\_spike[N][N]$, $prediction[N][N]$

\STATE Initialize parameters: 
\STATE \hspace{1 em} $fraction \gets 0.5$, $threshold \gets 0.5$
\STATE \hspace{1 em} $amplified\_fraction \gets 1.5$

\FOR {$i, j \in \{0, 1, ..., N-1\} \times \{0, 1, ..., N-1\}$}
    \STATE $(reduced\_spike[i][j], Vstore[i][j]) \gets$
    $\text{SpikeDivision}(DVS[i][j], fraction, threshold)$
\ENDFOR

\FOR {$i, j \in \{0, 1, ..., N-1\} \times \{0, 1, ..., N-1\}$}
    \STATE Compute neighborhood states: 
    \STATE \hspace{1 em} $top \gets (i_1 \geq 1) \ ? \ Vstore[i_1 - 1][j_1] \ : \ 0$
    \STATE \hspace{1 em} $bottom \gets (i_1 + 1 < N) \ ? \ Vstore[i_1 + 1][j_1] \ : \ 0$
    \STATE \hspace{1 em} $left \gets (j_1 \geq 1) \ ? \ Vstore[i_1][j_1 - 1] \ : \ 0$
    \STATE \hspace{1 em} $right \gets (j_1 + 1 < N) \ ? \ Vstore[i_1][j_1 + 1] \ : \ 0$
    \STATE \hspace{1 em} $amplified\_spike[i][j] \gets$
    \STATE \hspace{2 em} $\text{DigitalMS}(DVS[i][j], top, bottom, left, right, amplified\_fraction)$
\ENDFOR

\FOR {$i, j \in \{0, 1, ..., N-1\} \times \{0, 1, ..., N-1\}$}
    \STATE $spike\_occurred[i][j] \gets \mathbb{I}(DVS[i][j] > 0)$
\ENDFOR

\FOR {$i, j \in \{0, 1, ..., N-1\} \times \{0, 1, ..., N-1\}$}
    \STATE Compute 4-neighbor states: 
    \STATE \hspace{1 em} $top\_neighbors \gets \left[ \left( i \geq k+1 \right) ? Vstore[i-k][j] : 0 \, \middle| \, k \in \{0, 1, 2, 3\} \right]$

    \STATE \hspace{1 em} $bottom\_neighbors \gets \left[ \left( i + k < N \right) ? Vstore[i+k][j] : 0 \, \middle| \, k \in \{1, 2, 3, 4\} \right]$

    \STATE \hspace{1 em} $left\_neighbors \gets \left[ \left( j \geq k+1 \right) ? Vstore[i][j-k] : 0 \, \middle| \, k \in \{1, 2, 3, 4\} \right]$

    \STATE \hspace{1 em} $right\_neighbors \gets \left[ \left( j + k < N \right) ? Vstore[i][j+k] : 0 \, \middle| \, k \in \{1, 2, 3, 4\} \right]$
    
    \STATE \hspace{1 em} $prediction[i][j] \gets$  $\text{PredictNode}(top\_neighbors, bottom\_neighbors,$
    \STATE \hspace{1 em} $left\_neighbors, right\_neighbors, spike\_occurred[i][j])$
\ENDFOR

\RETURN $amplified\_spike, prediction$
\end{algorithmic}
\label{alg:motion_prediction}
\end{algorithm}

\subsection{Complete Digital Implementation of MP} \label{sec_dig_ckt}
We also introduce an alternative fully digital design approach through an automated digital flow, employing Questa Modelsim for simulation and Synopsys DC Compiler for synthesis using the GF 22nm FDSOI node. This approach is particularly suited for scenarios where power limitations are not the primary concern. Algorithm 6 details the implementation of a retina-inspired motion prediction circuit using the digital blocks. The design includes key motion prediction computational components, such as the voltage divider, predictive node, and amplified spike, all realized through digital blocks. 
Lines 4 to 6 implement a resistive divider equivalent using a floating-point multiplier, storing the value at $Vstore$ (\si{V_S}) once it surpasses the threshold. Lines 7 to 15 correspond to the Motion-Selective (MS) circuit mentioned earlier, responsible for generating the amplified spike for predicting object motion. Line 17 represents the \si{V_{SP}} node, which holds the value if a bipolar signal indicates that the object has passed that node. Lines 19 to 27 define our predictive node (\si{V_P}), which offers a four-level prediction, similar to having 16 transistors connected to the \si{V_P} node as in the mixed-signal approach. The mixed-signal readout circuit is implemented using an asynchronous AER scheme, while the digital flow necessitates the use of additional flip-flops to store intermediate results and outputs, as well as a threshold circuit.

\begin{figure}[b]
    \centering
    \fbox{\includegraphics[width=0.9\textwidth]{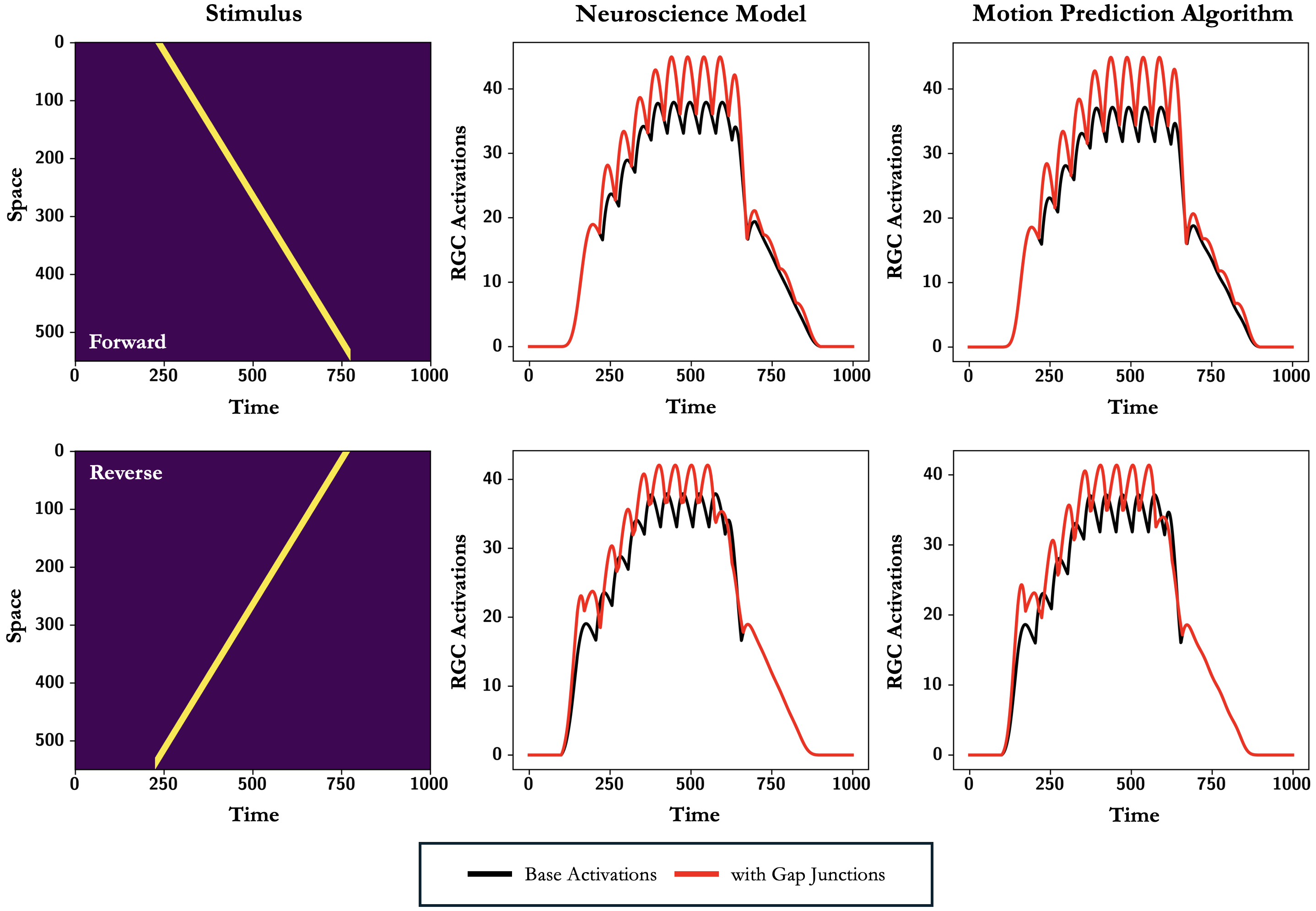}}
    \caption{1D experimental evaluation with our MP algorithm compared against a neuroscience model of motion prediction from \cite{SCHWARTZ2021246}.}
    \label{fig:mp-exp-1d}
\end{figure}

\section{Results \& Discussion} \label{sec_results}
This section presents the algorithmic and hardware verification results of our proposed retina-inspired motion prediction framework.

\subsection{Algorithmic Implementation}

\begin{figure}[b]
    \centering
    \fbox{\includegraphics[width=0.9\textwidth]{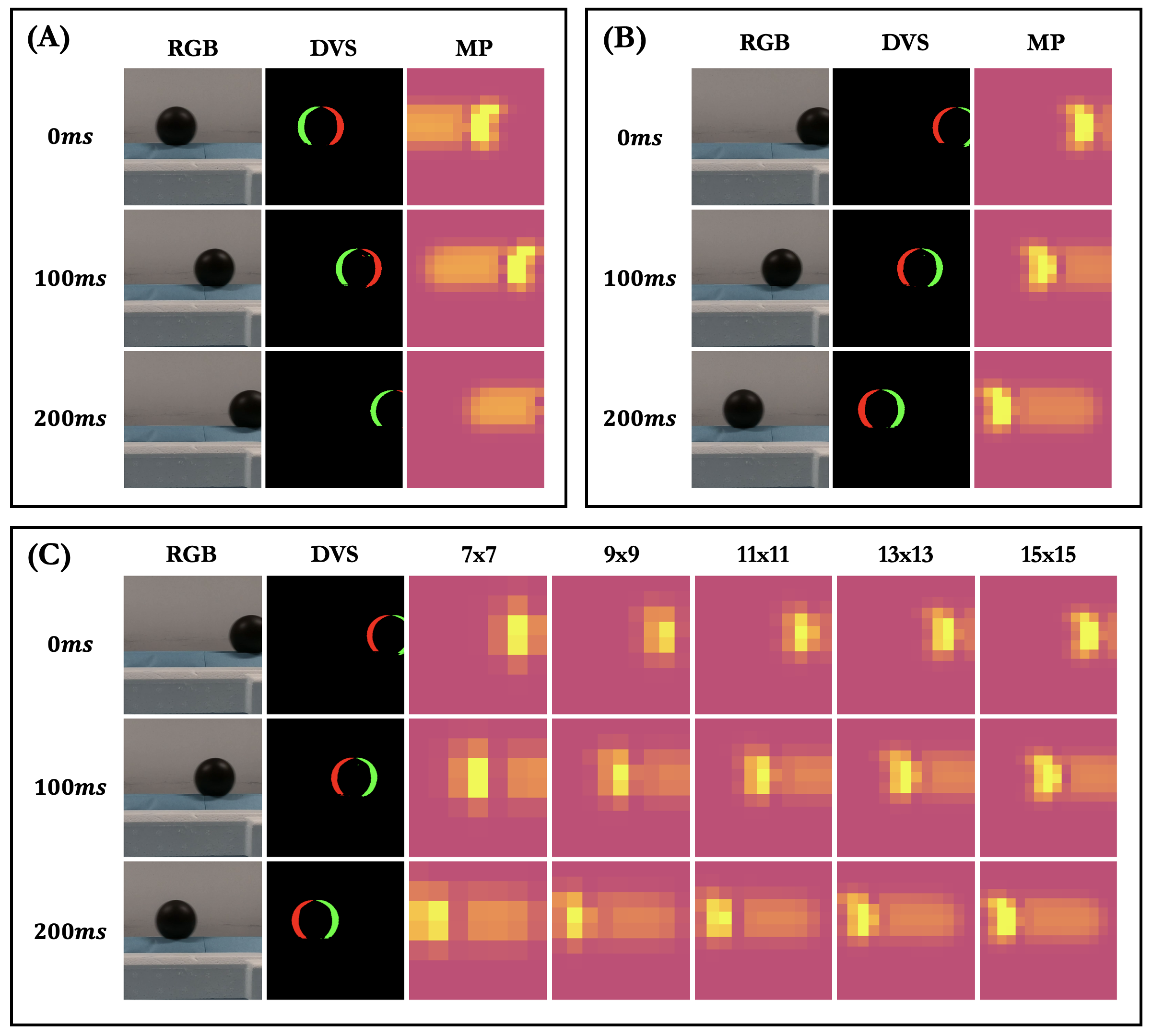}}
    \caption{Experimental evaluation with our MP algorithm. (A) Our MP algorithm at three intervals: 0 ms, 100 ms, and 200 ms with a sequence from \cite{QIN2024106502}. (B) Our MP algorithm with motion in the opposite direction of (A). (C) Evaluation of five motion prediction array sizes: 7\si{\times}7, 9\si{\times}9, 11\si{\times}11, 13\si{\times}13, and 15\si{\times}15.}
    \label{fig:mp-exp-2d}
\end{figure}

We evaluate the software implementation of our MP circuit using a simulated motion stimulus from \cite{SCHWARTZ2021246} and a real-world data sequence from \cite{QIN2024106502}, which we refer to as S1 and S2, respectively. S1 is a simulated sequence depicting one-dimensional object motion in both left and right (forward and backward) directions. S2, captured in a laboratory environment, features a black ball moving horizontally across a screen recorded with an RGB camera at 60 frames per second and a resolution of \si{480\times 720}. We pre-process this sequence in two steps: (1) cropping the images horizontally to ensure a square final resolution and (2) converting the images to DVS format through grayscale frame differencing, where pixels increasing in luminance generate a positive bipolar signal (green), while those decreasing in luminance generate a negative bipolar signal (red). To reduce noise in this process, we apply a spiking threshold of 50 to the individual pixel deltas before generating a spike.

\begin{figure}[b]
    \centering
    \fbox{\includegraphics[width=0.9\textwidth]{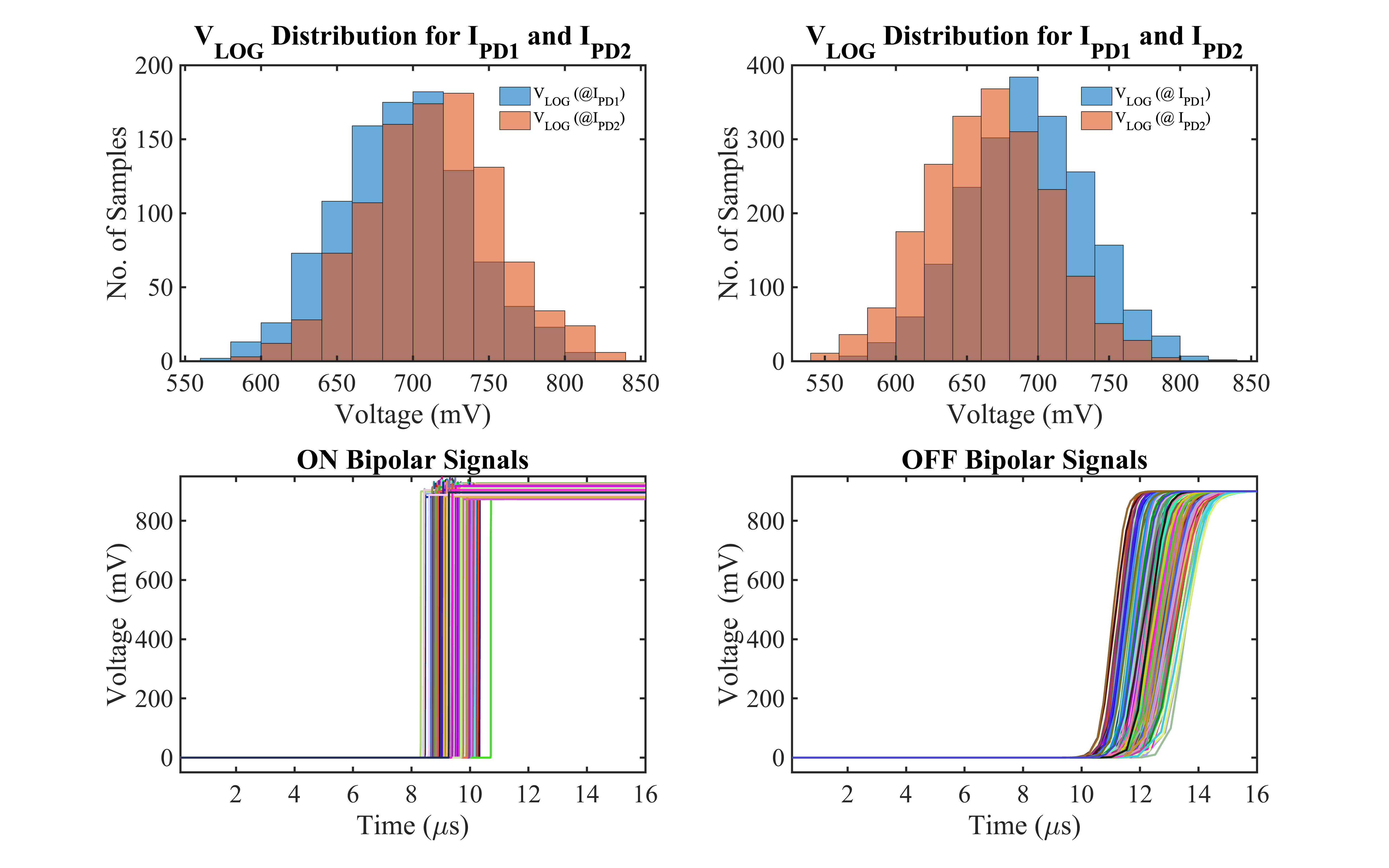}}
    \caption{Photoreceptor output voltage (\si{V_{LOG}} as shown in Figure \ref{fig:MP1}(a)) distribution for two different illuminances in DVS-based circuits, considering local mismatch and global supply voltage variations for both ON (left) and OFF (right) bipolar signals.}
    \label{fig:DVS}
\end{figure}

\begin{figure}[b]
    \centering
    \fbox{\includegraphics[width=0.9\textwidth]{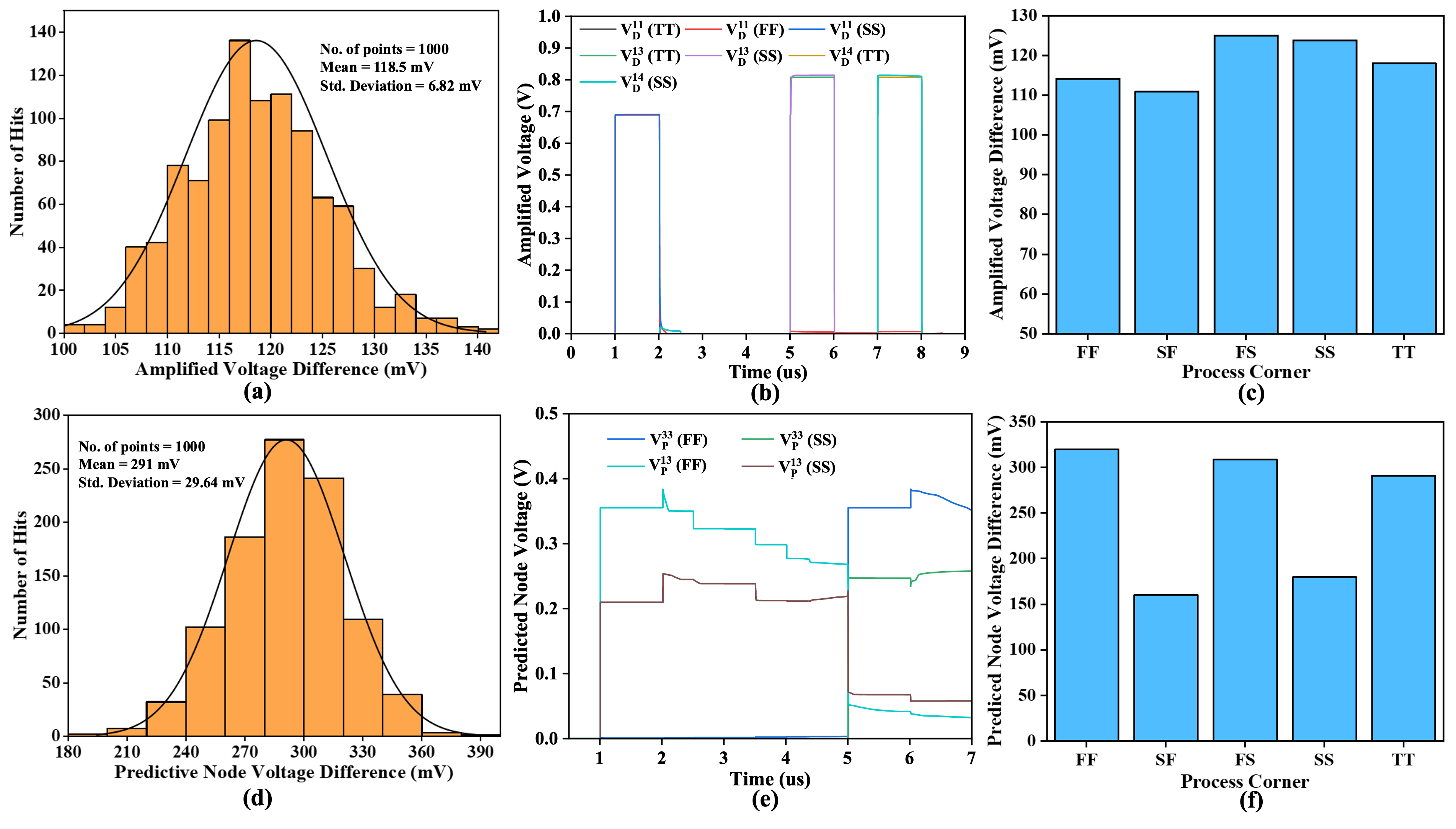}} 
    \caption{(a) Amplified Spike voltage on RGC node (\si{V_D}) difference distribution with respect to non-amplified spike considering local mismatch and global supply variation; (b) and (c) Variability analysis of amplified spike across process corners; (d) Predictive node voltage difference distribution with respect to non-predictive node voltage considering local mismatch and global supply variation; (e) and (f) Variability analysis of predictive node spike across process corners.}
    \label{fig:VAR_RES}
\end{figure}

In Figure \ref{fig:mp-exp-1d}, we compare our MP algorithm against the neuroscience model from \cite{SCHWARTZ2021246} in a one-dimensional setting to emphasize the increased RGC activity in the direction of predicted motion. These results indicate that our algorithm closely matches the neuroscience model in motion prediction while also generalizing to multiple directions. Furthermore, our implementation of gap junctions to connect multiple RGC cells achieves performance comparable to the neuroscience model, indicating that our approach closely replicates its biological counterpart by enhancing current transferability during motion prediction.

As shown in Figure~\ref{fig:mp-exp-2d}(A), we visualize the output of our MP algorithm for time-deltas of 100 ms using S2. This value was chosen based on recent findings in experimental neuroscience, which identify 100 ms as the forecast limit of the biological MP circuit \cite{SCHWARTZ2021246}. Our results demonstrate that the MP algorithm successfully anticipates future motion, as indicated by increased spiking activity in regions where motion is expected (Brighter regions - amplified spikes). The trail behind the bright regions signifies the suppression of predictive spikes due to the presence of a biphasic filter. To further validate the model, we reverse the direction of motion in Figure~\ref{fig:mp-exp-2d}(B), confirming the circuit’s ability to predict motion in both directions. Additionally, Figure \ref{fig:mp-exp-2d}(C) illustrates how varying the MP array size affects prediction characteristics: smaller arrays yield coarser predictions extending further into the future, while larger arrays provide higher-resolution predictions with a shorter temporal range. This trade-off between spatial resolution and temporal prediction will be crucial for downstream applications, and we plan to explore its implications in future work.

\subsection{Circuit Implementation}

We validate the motion prediction functionality of our proposed retina-inspired neuromorphic camera using HSPICE simulations with GlobalFoundries' 22nm FDSOI technology. To evaluate the DVS-based contrast detection circuit, simulations are conducted considering local mismatches and global supply voltage variations, incorporating a 10 mV standard deviation in the nominal supply voltage to account for supply voltage fluctuations. Figure \ref{fig:DVS} presents Monte Carlo (MC) simulation results based on over 1,000 samples for positive (left) and negative (right) contrast change scenarios. In these simulations, \si{I_{PD1}} and \si{I_{PD2}} represent photodetector currents at two different time instants, with \si{I_{PD1} > I_{PD2}}, at 30\% contrast sensitivity, which refers to the smallest luminance change between frames that triggers ON/OFF bipolar signals. In our extensive testing, the output voltage of each photoreceptor shows a standard deviation of 43.2 mV, causing overlapping voltage distributions ($\si{V_{LOG}}$) for \si{I_{PD1}} and \si{I_{PD2}} across various test samples. The worst-case difference in photoreceptor voltages between two consecutive frames among the 1,000 samples is 18.4 mV, based on 30\% contrast sensitivity. The difference voltage is amplified, and when it exceeds the threshold voltage, the DVS pixels asynchronously generate ON/OFF bipolar signals at different time instants, influenced by local variations and supply voltage fluctuations. The bipolar signal generation energy was 1.34 pJ, and the biasing energy was 0.32 pJ for a 1 \si{micro s} simulation.

The timing waveforms of the amplified spike on the RGC node (\si{V_D}) and the predictive node (\si{V_P}), obtained from the proposed MP circuit under various process corners, are shown in Figure \ref{fig:VAR_RES} (b) and (e). The amplified and predictive spikes of an MP array were evaluated in a controlled test setup, where an object moves in one of four directions: left to right, right to left, top to bottom, or bottom to top. The signals \si{V_D^{11}}, \si{V_D^{13}}, and \si{V_D^{14}} represent the amplified outputs, highlighting the enhancement achieved through the predictive nature of the gap junction. For simulation, the supply voltage (\si{Vdd}) was set to 0.7 V, while the amplification voltage for the MS (\si{Vddk}) was set to 0.9 V. A 16\si{\times}16 2D MP array was simulated over 1 \si{\micro s} per bipolar signal, with energy consumption measured at 18.56 pJ/event for four prediction levels in each direction, resulting in a total of 16 predictive cells. To evaluate the robustness of the MP circuit, we included both local mismatch effects and global supply voltage variations. The differences in spike levels between the predictive and amplified spikes, which characterize the MP behavior, were analyzed using Monte Carlo simulations with 1000 samples, as shown in Figure \ref{fig:VAR_RES} (a) and (d). The mean voltage difference for the amplified spikes was 118.5 mV with a standard deviation of 6.82 mV, while for the predictive spikes, the mean voltage difference was 291 mV with a standard deviation of 29.64 mV. As shown in Figure \ref{fig:VAR_RES} (c) and (f), the voltage difference remains stable despite process and mismatch variations, demonstrating the robustness of the MP circuit under different operating conditions.

\begin{table}[b]
\centering
\renewcommand{\arraystretch}{1}
\begin{tabular}{|l|c|c|c|}
\hline
\multicolumn{1}{|l|}{\small \textbf{Energy Type}} & \small \textbf{DVS Chip +} & \small \textbf{DVS Chip +} & \small \textbf{DVS Chip +} \\
\multicolumn{1}{|l|}{\small } & \small \textbf{Digital MP} & \small \textbf{Mixed Signal MP} & \small \textbf{Digital MP} \\
\multicolumn{1}{|l|}{\small } & \small \textbf{} & \small \textbf{(3D Integration)} & \small \textbf{(3D Integration)} \\
\hline
MP Compute Energy & 2.24 nJ & 0.519 nJ & 2.24 nJ \\
\hline
MP Readout Energy & 345.52 pJ & 4.9336 pJ & 4.9336 pJ \\
\hline
MP Total Energy & 2.585 nJ & 0.5239 nJ & 2.2449 nJ \\
\hline
Normalized MP Total Energy & 1 & 0.202 & 0.868 \\
\hline
\end{tabular}
\label{energy}
\caption{Energy consumption comparison for different configurations.}
\end{table}

\subsection{Energy Comparison}

Table 2 presents the energy consumption for three proposed configurations: (a) a DVS sensor chip with a digitally implemented motion prediction logic chip integrated on the same PCB, (b) a DVS sensor chip with a 3D integrated mixed-signal motion prediction circuit, and (c) a DVS sensor chip with a 3D integrated digital motion prediction chip. The first approach, which combines the DVS chip and the digital motion prediction (MP) on the same PCB, accounts for energy used in DVS static and event generation, as well as the digital motion prediction circuit and the low-voltage differential signaling (LVDS) utilized for data transfer between the sensor and the MP die. The second and third architectures utilize 3D integration, employing either Cu-Cu hybrid bonding or through-silicon vias (TSVs), in addition to the DVS static and event generation energy. The second architecture adopts a mixed-signal approach, while the third architecture uses a purely digital design. The table includes a simulation scenario with a \si{7\times 7} MP array, which generates 28 spikes, each lasting 1 \si{\micro s}, over a 1 ms simulation period. The total energy consumption for DVS static and event generation is 15.68 nJ and 37.52 pJ, respectively. The energy consumption for LVDS is 12.34 pJ/bit \cite{lvds_energy}, while 3D integrated die-to-die communication consumes only 176.2 fJ/bit \cite{tsv_energy}. From the table, it can be observed that the 3D integrated mixed-signal implementation offers significantly better energy efficiency compared to other two configurations.

 \begin{table*}[t] 
    \centering
    \scalebox{0.75}{
    \begin{tabular}{|l|c|c|c|c|c|}
        \hline
        & \textbf{2013 \cite{6572366}} & \textbf{2021 \cite{9540043}} & \textbf{2023 \cite{phillip2025neuromorphicretinafpgabasedemulator}} & \textbf{2024 \cite{peng2024multifunctional}} & \textbf{This work} \\
        \hline
        \textbf{Material} & CMOS & CMOS & CMOS & WS2 & \textbf{22nm GF FDSOI} \\
        \hline
        \textbf{Model} & Motion Tracking & General & General & Motion Tracking & \textbf{Motion Prediction} \\
        \hline
        \textbf{Functionality} & Temporal & Temporal & Spatio-Temporal & Spatio-Temporal & \textbf{Spatio-Temporal} \\
        \hline
        \textbf{Mixed Signal Flow} & Yes & No & No & Yes & \textbf{Yes} \\
        \hline
        \textbf{Complete Digital Flow} & No & Yes & Yes & No & \textbf{Yes} \\
        \hline
        \textbf{Motion Tracking} & Yes & No & No & Yes & \textbf{Yes} \\
        \hline
        \textbf{Motion Prediction} & No & No & No & No & \textbf{Yes} \\
        \hline
        \textbf{Ganglion Cell Type} & NA & No & ON and OFF & ON and OFF & \textbf{ON and OFF} \\
        \hline
        \textbf{Speed Adaption} & No & No & No & Yes & \textbf{Yes} \\
        \hline
    \end{tabular}
    }
    \caption{Comparison with previous state-of-the art work}
    \label{tab:comparison}
\end{table*}

\subsection{Comparison with previous works}
To the best of our knowledge, no prior work has demonstrated on-chip MP in real time. This work represents the first implementation of real-time MP, with potential applications in threat detection and autonomous navigation. Table 3 provides a summary comparing our proposed retina-inspired framework with previous studies. Previous studies~\cite{peng2024multifunctional, 6572366} have primarily focused on motion tracking or direction prediction, both of which are inherently supported by our proposed architecture. We introduce a novel MP hardware design capable of predicting surrounding pixel motion by leveraging computational principles inspired by the biological retina. The proposed system is implemented using an advanced CMOS foundry-supported process design kit (PDK), which requires no specialized materials for fabrication. Furthermore, we present a digital equivalent model of MP, which can be deployed on an FPGA for specific applications rather than general retinal emulation. This model incorporates the leaky integrate-and-fire (LIF) neuron model on an FPGA, as previously demonstrated in \cite{phillip2025neuromorphicretinafpgabasedemulator, 9540043}. Additionally, we propose a 3D-integrated mixed-signal architecture that optimizes both energy efficiency and area utilization, offering a scalable and high-performance solution for real-time MP.

\section{Conclusion} \label{sec_conclusion}
Our proposed retina-inspired neuromorphic framework bridges advancements in retinal neuroscience, mixed-signal circuit techniques, and algorithmic modeling to achieve efficient, pixel-level motion prediction (MP) for machine vision applications. By identifying and selecting only the most essential retinal features necessary for MP, we reduce design complexity, minimize energy consumption, and lower area overhead, making our solution well-suited for integration into CMOS image sensors. Our hardware-algorithm framework is implemented using GlobalFoundries’ 22nm FDSOI technology and offers two design approaches: a mixed-signal implementation that emulates key biological retinal functions using specialized computational elements—such as a biphasic filter, signal divider, spike adder, and non-linear circuit—and a fully digital implementation that serves as a comparison for our proposed circuit.  Furthermore, we introduce a novel 3D integration approach using Cu-Cu hybrid bonding with one hybrid bond per pixel, significantly optimizing area utilization and simplifying routing complexity, enabling a more compact and efficient hardware design. Validated on real-world object stimuli, our model demonstrates its ability to perform accurate MP with low energy consumption, achieving just 18.56 pJ/MP in our mixed-signal implementation. These results highlight the potential of our neuromorphic framework in predictive visual computation for decision-making applications, including robotics, autonomous navigation, and augmented reality. 

\section*{Acknowledgment}
The research was funded in part by National Science Foundation through awards CCF2319617 and CCF2319619.

\section*{References}
\bibliographystyle{ieeetr}
\bibliography{references}

\end{document}